\documentclass [12pt] {article}
\usepackage{amsmath,amssymb,amsfonts,amsthm}
\usepackage{multirow}
\usepackage{lscape}
\usepackage{color}
\newcommand{\cs}[3]{{{#3} \brace {#1 #2}}}

\newcommand{\h}[1]{\mathop{\hat{\lambda}}\limits_{#1}\ \!\!\!}
\newcommand{\q}[1]{\mathop{\lambda}\limits_{#1}\ \!\!\!}
\newcommand{\C}[1]{\mathop{C}\limits_{#1}\ \!\!\!}

\newcommand{\edf}{\ {\mathop{=}\limits^{\rm def.}}\ }

\newcommand{\E}[1]{\mathop{L}\limits_{#1}\ \!\!\!}
\begin{document}

\begin{center}
{ \large \bf An AP-Structure With Finslerian Flavor \\
II: Torsion, Curvature and Other Objects}\\
\end{center}
\begin{center}
\bf{M.I.Wanas\footnote{Astronomy Department, Faculty of Science,
Cairo University, Giza, Egypt.

 E-mail:mwanas@cu.edu.eg; wanas@frcu.eun.eg}}$^{,3,4}$ and \bf{Mona M. Kamal\footnote{Mathematics Department, Faculty of Girls,
Ain Shams University, Cairo, Egypt.

E-mail:monamkamal@eun.eg}}$^{,}$\footnote{Egyptian Relativity Group (ERG) URL:http://www.erg.eg.net

~$^{4}$Center for Theoretical Physics (CTP) at the British University in Egypt (BUE).}

\end{center}
\begin{abstract}
 An absolute parallelism (AP-) space having Finslerian properties is called FAP-space. This FAP-structure is more wider than both conventional AP and Finsler structures. In the present work, more geometric objects as curvature and torsion tensors are derived in the context of this structure. Also second order tensors, usually needed for physical applications, are derived and studied. Furthermore, the anti-curvature and the W-tensor are defined for the FAP-structure. Relations between Riemannian, AP, Finsler and FAP structures are given. These relations facilitate comparison between results of applications carried out in the framework of these structures. We hope that the use of the FAP-structure, in applications may throw some light on some of the problems facing geometric field theories.
\end{abstract}
\section{Introduction and Motivations}

Recently, a lot of attention has been paid to Finsler geometry for applications (cf. Ref. 
1, 2, 3). On the other hand,
teleparallel geometry (absolute parallelism (AP-) geometry) has gained many authors 
(cf. Ref. 4, 5, 6)
for the purpose of physical applications. The interest in such geometries has been grown for the goal of solving
problems of gravity theories, recently discussed (cf. Ref. 7, 8, 9). The aim has been to construct field theories,
including general relativity (GR), in more wider geometries than the Riemannian one.

In this direction, one of the authors (MIW) has  attempted to find a type of geometry with both AP and Finsler properties
(FAP-structure). The geometric structure  proposed \cite{W1} is more wider than both AP and Finsler geometries.
The present work is the second part of this work, in which torsion, curvature and other objects
are derived in the FAP. The motivation of this work is given, in some details, in the first part  \cite{W1}.

The paper is organized in the following manner. In the next Section, we give a brief review of the results obtained
in the first part. Second order tensors and scalars, that may be important for
physical applications, are given in Section 3. In Section 4, torsion, curvature and anti-curvature tensors, of the FAP-structure, are derived. In Section 5,
the W-tensor is derived for the present structure. The paper is discussed and concluded in Section 6. In what follows we are going to characterize geometric objects, in the FAP-structure, by putting a hat on the same notations used in AP-geometry.

\section{Summary of The Results of the First Part}
In first part  \cite{W1} the following results have been obtained: \\

{\bf{Definition}:} An Absolute Parallelism space with  Finslerian (FAP)
properties $(~M, \E{i}~)$ is an $n$-dimensional differentiable manifold
$M$ equipped with a set of n-Lagrangian functions $\E{i}\equiv
\E{i}(x,y)$ having the
following properties: \\ \\
1.  $\E{i}(x,y)$ is $C^{\infty}$ on $\tau M (= TM \backslash \{0 \})$.  \\
2. $\E{i}(x,y)> 0, y~ \in~ \tau M , ~ ~ y=\dot{x}~(=\frac{dx}{dt}) $.
\\ 3. $\E{i}(x,y)$ is positively homogenous of degree one  ($P-h(1)$).  \\
4. The vector fields  \footnote{In what follows the colon (:) is used for differentiation w.r.t. y, while comma (,) is used for differentiation w.r.t. x., $\mu(=1, 2, ..., n)$ represents  the vector component and $i(=1, 2, ..., n)$ is the vector number. }
\begin{equation}\label{1}
\h{i}_{\mu}  (x,y) \edf  \frac{\partial \E{i}}{\partial y^{\mu}}=\E{i}_{: \mu},
\end{equation}
are assumed to be linearly independent and globally defined on $M$, then we can define the conjugate vector fields
 $\h{i}^{\mu}$ such that,
\begin{equation}\label{2}
\h{i}^{\mu}\h{i}_{\nu} = \delta^{\mu}_{\nu}, ~~ \end{equation}
\begin{equation}\label{3}
\h{i}^{\mu}\h{j}_{\mu} = \delta_{ij}.
\end{equation}
These vector fields are the \underline{building blocks} of the FAP-Structure. The following object,
\begin{equation}\label{4}
N^{\nu}_{. \alpha} \edf y^{\mu} \h{i}^{\nu} \h{i}_{\mu, \alpha},
\end{equation}
is found to transform as a {\it non-linear connection}.
 The following second order tensor,
 \begin{equation}\label{5}
 \C{i}_{\mu \alpha} \edf \frac{\partial
\h{i}_{\mu}}{\partial y^{\alpha}}=\E{i} _{ :\mu  \alpha } ,
\end{equation}
plays an important role in the FAP-Structure. It is type (0,2) symmetric tensor and P-h(-1).
The following theorem has been  proved.\\

{\bf {{Theorem 1}.}} {\it "A necessary and sufficient condition for an
FAP-space to be an AP-space is that $\C{i}_{\alpha \beta}$  vanishes
identically"}.\\ \\
It is shown that, if
\begin{equation}\label{6}
\hat{C}^{\mu}_{.~ \alpha \beta } \edf \h{i}^{\mu} \C{i}_{\alpha \beta}, \end{equation}
then,
\begin{equation}\label{7}
\C{i}_{\alpha \beta} = \h{i}_{\mu}~\hat{C}^{\mu}_{.~ \alpha \beta }. \end{equation}
Using the non-linear connection (\ref{4}), we can define the following operator
\begin{equation}\label{8}\delta_{\mu} \edf \partial_{\mu} - N^{\alpha}_{.\mu}
\frac{\partial}{\partial y^{\alpha}}. \end{equation}
This operator can be written as $(\hat{,})$, if it is used as an infix operator.\\

The following geometric objects have been shown to transform as  linear connections in the FAP-space:

\begin{enumerate}
  \item {\bf {\it Cartan-Like} Linear Connection},
  \begin{equation}\label{9}\hat{\Gamma}^{\mu}_{.~ \alpha \beta} \edf {\h{i}}^{\mu} \h{i}_{\alpha
\hat{,} \beta}. \end{equation}

 Since (\ref{9}) is non-symmetric, we can define the dual linear connection,
    \begin{equation}\label{28}\widetilde{\hat{\Gamma}} ~^{\mu}_{.~ \alpha \beta} \edf \hat{\Gamma}^{\mu}_{.~\beta \alpha }, \end{equation}
   and also the  symmetric linear connection\footnote{The parentheses $(~)$ stands for the symmetric part  while the brackets $[~]$ stands for the skew part, for any geometric object, w.r.t. the indices included.},
  \begin{equation}\label{29}\hat{\Gamma}^{\mu}_{.~ (\alpha \beta)} \edf \frac{1}{2}(\hat{\Gamma}^{\mu}_{.~ \alpha \beta}+\hat{\Gamma}^{\mu}_{.~\beta \alpha }). \end{equation}

  \item {\bf {\it Bervald-Like} Linear Connection},
  \begin{equation}\label{10}
  \hat{B}^{\mu}_{.~ \alpha \beta} \edf N^{\mu}_{~\alpha : \beta} =
\frac{\partial}{\partial y^{\beta}} (y^{\nu} \h{i}^{\mu}\h{i}_{\nu,~
\alpha}).\end{equation}
\end{enumerate}
Connections (\ref{9}), (\ref{28}) and (\ref{10}) are non-symmetric in their lower indices $(\alpha, \beta)$. Consequently, they possess torsion, respectively.\\
Using the objects (\ref{9}), (\ref{28}), (\ref{29}) and (\ref{6}) one can define the following covariant derivatives, respectively:\\\\
If $A_\mu$ is an arbitrary vector, then its \underline{horizontal (H-) positive derivative} is defined by
\begin{equation}\label{11}
A_{\stackrel {\alpha||\beta}{+~~} }\edf A_{\alpha \hat{,} \beta} - A_{\mu}
\hat{\Gamma}^{\mu}_{. \alpha \beta}.\end{equation}
Also, we can define the following \underline{H-negative derivative} using the dual connection, as follows
\begin{equation}\label{30}
A_{\stackrel{\alpha|| \beta}{-~~} } \edf A_{\alpha \hat{,} \beta} - A_{\mu}
\widetilde{\hat{\Gamma}}~^{\mu}_{. \alpha \beta}.\end{equation}
In addition we can define a third \underline{H-neutral derivative} using the symmetric part of the {\it Cartan like} linear connection, i.e.
\begin{equation}\label{31}
A_{\alpha || \beta} \edf A_{\alpha \hat{,} \beta} - A_{\mu}
\hat{\Gamma}^{\mu}_{. (\alpha \beta)}.\end{equation}

The  \underline{vertical (V-) derivative} of the vector $A_\mu$, is defined by
\begin{equation}\label{12}A_{\mu}|_{\nu} \edf A_{\mu : \nu} - A_{\alpha}\hat{C}^{\alpha}_{.\mu
\nu}.\end{equation}
Since $\hat{C}^{\alpha}_{.\mu\nu}$ is symmetric  in its two  lower indices $(\mu, \nu)$, then we have only one type of derivative.\\
It has been shown that,
\begin{equation}\label{13}
\h{i}_{\mu}|_{\nu}  \equiv 0,\end{equation}
\begin{equation}\label{14}
\h{i}_{\stackrel {\mu||\nu}{+~~} } \equiv 0.  \end{equation}
Relations (\ref{13}) and (\ref{14}) show clearly that $\h{i}_\mu$ are parallel displaced along certain paths characterized by the geometric objects (\ref{6}) and (\ref{9}), respectively.\\
The second order symmetric tensors,
\begin{equation}\label{15}
\hat{g}_{\mu \nu} \edf  \h{i}_{\mu} \h{i}_{\nu} ,  \end{equation}
\begin{equation}\label{16}
\hat{g}^{\alpha \beta}  \edf  \h{i}^{\alpha} \h{i}^{\beta},  \end{equation}
can be used to play the role of the metric tensor (and its conjugate) of Finsler space associated with the FAP-space, under certain condition \cite{W1}.
It is clear from the definition (\ref{15}) 
and the identities (\ref{14}) that
\begin{equation}\label{17}
\hat{g}_{\stackrel {\mu \nu|| \sigma}{++~~~} } \equiv0,\end{equation}
which means that the linear connection (\ref{9}) is a metric one. Also, it is shown that
\begin{equation}\label{18}
\hat{g}_{\mu \nu}|_{\sigma} \equiv 0.\end{equation}
Relations (\ref{17}), (\ref{18}) means simply that the operation of raising and lowering tensor indices commutes with the operations of H-positive and V- differentiation.\\ \\
{\bf  Further Relations}\\ \\
If we define the tensor,
 \begin{equation}\label{25}
 \stackrel {\star}{\C{i}}~^\mu_{.~\alpha} \edf \h{i}^{\mu}_{.~:\alpha} ,
\end{equation}
then we can write,
 \begin{equation}\label{26}
\stackrel {\star}{C}~^\mu_{.~\nu\alpha}\edf \h{i}^{\mu}_{.~:\nu} \h{i}_\alpha,
\end{equation}
which in virtue of (\ref{2}) and definition (\ref{6}) can be written as,

$$\stackrel {\star}{C}~^\mu_{.~\nu\alpha}= - ~\hat{C}^\mu_{.~\nu\alpha}.$$
Now, since $\hat{C}^\mu_{.~\nu\alpha}$ is symmetric w.r.t. its two lower indices, then we get,
 $$\stackrel {\star}{C}~^\mu_{.~[\nu\alpha]}=~0,$$
which gives immediately the relation,
\begin{equation}\label{27}
 \h{i}^{\mu}_{.~:\alpha} \h{i}_\nu =\h{i}^{\mu}_{.~:\nu} \h{i}_\alpha.
\end{equation}
This is an important relation characterizing the FAP-structure.\\

The canonical linear  connection (\ref{9}) can be expressed in the form
\begin{equation}\label{22}
\hat{\Gamma}^{\alpha}_{.~ \mu \nu}\edf\hat{\cs{\mu}{\nu}{\alpha}}+\hat{\gamma}^\alpha_{.\mu\nu},
\end{equation}
where  $\hat{\cs{\mu}{\nu}{\alpha}}$  is the  linear metric connection defined by,
$$\hat{\cs{\mu}{\nu}{\alpha}}\edf\frac{1}{2}\hat{g}^{\alpha \beta}
(\hat{g}_{\beta\mu \hat{,} \nu}+ \hat{g}_{\beta\nu\hat{,} \mu}-\hat{g}_{\mu \nu\hat{,} \beta}),
$$
and $\hat{\gamma}^\alpha_{.\mu\nu}$ is a third order tensor defined by,
\begin{equation}\label{0}
\hat{\gamma}^\alpha_{.\mu\nu}\edf\h{i}^\alpha\h{i}_{\mu\hat{;}\nu},
\end{equation}
the infix operator $(\hat{;})$ stands for the covariant differential operator using $\hat{\cs{\mu}{\nu}{\alpha}}$
i.e.\begin{equation}\label{-2}
\h{i}_{\mu\hat{;}\nu}\edf\h{i}_{\mu\hat{,}\nu}-\h{i}_\epsilon\hat{\cs{\mu}{\nu}{\epsilon}}\end{equation}

For the FAP-structure we have a set of, at least, four d-connections: $D=(\hat{\cs{\mu}{\nu}{\alpha}}, N^\alpha_{~\mu}$, $\hat{C}^\alpha_{.\mu\nu})$, $ \hat{D}=(\hat{\Gamma}^\alpha_{.\mu\nu},N^\alpha_{~\mu}, \hat{C}^\alpha_{.\mu\nu})$, $
\widetilde{D}=(\widetilde{\hat{\Gamma}}~^\alpha_{.\mu\nu}, N^\alpha_{~\mu}, \hat{C}^\alpha_{.\mu\nu}) $ and
$\overline{D}=({\hat{\Gamma}}~^\alpha_{.(\mu\nu)}, N^\alpha_{~\mu}, \hat{C}^\alpha_{.\mu\nu}).$
\section{Tensors of Different Orders}
In geometric field theories, different geometric objects play important roles in representing physical quantities,
in the context of the geometrization  philosophy. To describe different physical fields, one may need scalars, vectors and /or second order tensors. On the other hand, third order objects (connections or tensors) are necessary to describe trajectories of test particles in different physical fields. For these reasons, we are going to give definitions of such objects in the FAP-Structure. This is done in  order to make this structure ready for applications. Of course, in what follows we cannot cover all such objects, but only those that we expect, from previous experience, to play a role in physical applications.

\subsection{Third Order Tensors and Vectors}
Now, we are going to  give definitions for some important third order tensors and their contractions.
The torsion of the linear connection (\ref{9}) is given by,
\begin{equation}\label{38}
\hat{\Lambda}^\alpha_{.
\mu\nu}\edf\hat{\Gamma}^{\alpha}_{.~ \mu \nu}-\hat{\Gamma}^{\alpha}_{.~\nu \mu }=\hat{\gamma}^{\alpha}_{.~ \mu \nu}-\hat{\gamma}^{\alpha}_{.~\nu \mu },\end{equation}
which gives by contraction the vector,
\begin{equation}\label{39} \hat{c}_\alpha\edf\hat{\Lambda}^\mu_{.\alpha\mu}=\hat{\gamma}^\mu_{.\alpha\mu}.\end{equation}
Also, the torsion of the {\it Bervard-like} connection (\ref{10}) is defined by,
\begin{equation}\label{42} \hat{T}^\alpha_{.
\mu\nu}\edf\hat{B}^{\alpha}_{.~ \mu \nu}-\hat{B}^{\alpha}_{.~\nu \mu },\end{equation}
which gives the vector,
\begin{equation}\label{43} \hat{T}_{
\mu}\edf\hat{T}^\alpha_{.
\mu\alpha}.\end{equation}
Now, using the contortion (\ref{0}), we can define the third order tensors,
\begin{equation}\label{45}
\hat{\Delta}^\alpha_{.\mu\nu}\edf\hat{\gamma}^\alpha_{.\mu\nu}+\hat{\gamma}^\alpha_{.\nu\mu},\end{equation}
\begin{equation}\label{46}
\hat{\gamma}^{~~\alpha}_{\mu\nu.}~\edf \hat{g}^{\epsilon \alpha} \hat{g}_{\mu\beta} \hat{\gamma}^\beta_{.\nu\epsilon}\end{equation}
Also, using the non-linear connection (\ref{4}) and the operator (\ref{8}), we can define the object,
\begin{equation}\label{40} N ^\alpha_{.\mu\nu}\edf N^\alpha_{.~\mu\hat{,}\nu} - N^\alpha_{.~\nu\hat{,}\mu}.\end{equation}
This object can be shown (see Sec. 4) to be tensor of type (1,2). Its contraction  gives the vector,
\begin{equation}\label{41} N_\mu\edf N^\alpha_{.~\mu\alpha}.\end{equation}
The contraction of the tensor (\ref{6}) will give the vector,
\begin{equation}\label{44}
\hat{V}_\alpha\edf\hat{C}^\mu_{.\alpha\mu}\end{equation}

\subsection{Second Order Tensors and Scalars}

In the AP-geometry, a set of second order tensors has been defined \cite{M}. Such tensors have been used to classify \cite{MW},
physically, different AP-structures  (with different symmetries) used for applications. This classification
scheme   is a covariant one  \cite{M0}. It gives the capabilities of  certain AP-structures to represent
physical situations. Also, some physical meaning have been attributed to these tensors. Such tensors
are obtained using the torsion, contortion tensors of the linear connection and their contraction.
So, it would be interesting to define the corresponding second order tensors in the context
of the FAP-structure. This would facilitate comparison with the physical results obtained
from Ap-geometry, when using FAP-geometry for physical applications.\\

The following Table gives a summary of second order tensors defined in the context of the FAP-geometry .
As stated above, we use the same notation of the AP-geometry, with a hat characterizing the FAP-geometry.
The condition for removing this hat is given by Theorem 1. Table 1 is similar to that of the AP-structure \cite{M}.

\begin{center}
{\bf Table 1: Second Order World Tensors}\\[0.2 cm]

\begin{tabular}{|l|l|}
   \hline
  Skew Tensors& Symmetric Tensors\\ \hline \hline
  $\hat{\xi}_{\mu\nu}\edf\hat{\gamma}_{\stackrel {\mu\nu.||\epsilon}{~~~~+}}^{~~\epsilon}$&  \\
 $\hat{\zeta}_{\mu\nu}\edf\hat{c_\alpha}\hat{\gamma}_{\mu\nu.}^{~~\alpha}$ &  \\ \hline
  $\hat{\eta }_{\mu\nu}\edf\hat{c_\alpha}\hat{\Lambda}^\alpha_{.\mu\nu}$&$\hat{ \phi}_{\mu\nu}\edf\hat{c}_\alpha\hat{\Delta}^\alpha_{.\mu\nu}$\\\hline
  $\hat{\chi }_{\mu\nu}\edf\hat{\Lambda}^\epsilon_{\stackrel{.\mu\nu||\epsilon}{~~~~+}}$& $\hat{\psi }_{\mu\nu}\edf\hat{\Delta}^\epsilon_{\stackrel{.\mu\nu||\epsilon}{~~~~+}}$\\\hline
  $\hat{\varepsilon }_{\mu\nu}\edf\hat{c}_{\stackrel{\mu||\nu}{+~~}}-\hat{c}_{\stackrel{\nu||\mu}{+~~}}$&$\hat{ \theta}_{\mu\nu}\edf\hat{c}_{\stackrel{\mu||\nu}{+~~}}+\hat{c}_{\stackrel{\nu||\mu}{+~~}}$ \\\hline
  $\hat{\kappa}_{\mu\nu}\edf\hat{\gamma}^\alpha_{.\mu\epsilon}\hat{\gamma}^\epsilon_{.\alpha\nu}-\hat{\gamma}^\alpha_{.\nu\epsilon}\hat{\gamma}^\epsilon_{.\alpha\mu}$&$\hat{\varpi}_{\mu\nu}\edf\hat{\gamma}^\alpha_{.\mu\epsilon}\hat{\gamma}^\epsilon_{.\alpha\nu}+\hat{\gamma}^\alpha_{.\nu\epsilon}\hat{\gamma}^\epsilon_{.\alpha\mu}$\\\hline
   & $\hat{\omega }_{\mu\nu}\edf\hat{\gamma}^\epsilon_{.\mu\alpha}\hat{\gamma}^\alpha_{.\nu\epsilon}$\\
   &$ \hat{\sigma }_{\mu\nu}\edf\hat{\gamma}^\epsilon_{.\alpha\mu}\hat{\gamma}^\alpha_{.\epsilon\nu}$\\
   & $\hat{\alpha}_{\mu\nu}\edf\hat{c}_{\mu}\hat{c}_{\nu}$ \\ \hline
\end{tabular}
\end{center}
where $ \hat{\gamma}_{\stackrel {\mu\nu.||\epsilon}{~~~~+}}^{~~\epsilon}\edf\hat{\gamma}_{\stackrel {\mu\nu~.~||\epsilon}{++~~~~}}^{~~~\stackrel\epsilon{+}}$ and so on. \\
We can obtain another set of second order tensors, resulting from the combination of (\ref{6}), (\ref{38}), (\ref{42}), (\ref{40}) and their contractions, (e.g.
skew-tensors,
($ N_\alpha\hat{\gamma}_{\mu\nu.}^{~~\alpha},~ N_\alpha\hat{\Lambda}^\alpha_{.\mu\nu},~
 \hat{c_\alpha}\hat{T}^\alpha_{.\mu\nu},~  N_\alpha N^\alpha_{.\mu\nu},~ ...$),
and symmetric-tensors,
($\hat{V_\alpha}\hat{V_\beta}, ~\hat{c_\alpha}\hat{C}^\alpha_{.\mu\nu},~ \hat{T}_\alpha \hat{C}^\alpha_{.\mu\nu},~\hat{T}_\mu\hat{T}_\nu,~...$)).
The properties of such tensors will be studied in details, after application, depending on their importance for such applications.\\

From the tensors given in Table 1, we can define many scalars (e.g. $\hat{ \phi}\edf\hat{c}_\alpha\hat{\Delta}^{\alpha\mu}_{..~\mu}$,
$\hat{\psi }\edf\hat{\Delta}^{\epsilon\mu}_{\stackrel{..~\mu||\epsilon}{~~~~+}}$,
$\hat{ \theta}\edf2~\hat{c}^\mu_{\stackrel{~||\mu}{~~+}}$,
$\hat{\alpha}\edf\hat{c}^{\mu}~\hat{c}_{\mu}$,
$\hat{\varpi}\edf\hat{\gamma}^{\alpha\mu}_{..~\epsilon}\hat{\gamma}^\epsilon_{.\alpha\mu}+\hat{\gamma}^\alpha_{.\mu\epsilon}\hat{\gamma}^{\epsilon.\mu}_{~\alpha}$,
$\hat{\omega }\edf\hat{\gamma}^{\epsilon\mu}_{..~\alpha}\hat{\gamma}^\alpha_{.\mu\epsilon}$ and $ \hat{\sigma }\edf\hat{\gamma}^\epsilon_{.\alpha\mu}\hat{\gamma}^{\alpha.\mu}_{~\epsilon}$).

Among these tensors, we have a set of second order traceless tensors (e.g. ($\hat{\phi}_{\mu\nu}+\hat{\alpha}_{\mu\nu}$), ($\hat{\psi}_{\mu\nu}+\hat{\theta}_{\mu\nu}$) and  ($\hat{\varpi}_{\mu\nu}+2~\hat{\omega}_{\mu\nu}$)) with possible linear combination. Such object may be of physical interest.


\section{Curvature, Anti-Curvature and Torsion Tensors}
It is well known that any differentiable manifold equipped with a d-connection,  $F=(F^{\alpha}_{.~ \mu \nu}, G^\alpha_{.\beta}, C^\alpha_{.\mu \nu})$,
implies the following commutation relations (cf. Ref. 14) 
$$A_{\alpha||\beta\sigma}-A_{\alpha||\sigma\beta}=A_\mu R^\mu_{.\alpha\beta\sigma}-A_\alpha|_{\mu}(G^\mu_{.\beta\hat{,}\sigma} - G^\mu_{.\sigma\hat{,}\beta})-A_{\alpha||\mu}\Lambda^\mu_{.\beta\sigma}
,
$$
$$
A_\alpha|_{\beta\sigma}-A_\alpha|_{\sigma\beta}=A_\mu S^\mu_{.\alpha\beta\sigma}-A_{\alpha}|_{\mu}(C^\mu_{.\beta\sigma}-C^\mu_{.\sigma\beta}),
~~~~~~~~~~~~~~~~~$$
where $A_\mu$ is an arbitrary vector field, function of both $x$ and $y$,  $C^\alpha_{.\mu \nu}$ is $3^{rd}$ order tensor, $G^\alpha_{.\beta}$ is a non-linear connection, $R^\mu_{.\alpha\beta\sigma}$ is the curvature using the linear connection $F^{\alpha}_{.~ \mu \nu}$ and $\Lambda^\mu_{.\beta\sigma}$ is the torsion of this connection. Also, $_{||}$ and $|$ denote  differentiation using $F^{\alpha}_{.~ \mu \nu}$ and $C^\alpha_{.\mu \nu}$, respectively. \\

Now using distinguish connections, $\hat{D}$, $\widetilde{D}$, $\overline{D}$ and $D$ of the FAP-structure, we can write the following horizontal commutation relations, respectively,

\begin{equation}\label{23}
A_{\stackrel{\alpha||\beta\sigma}{+~~~}}-A_{\stackrel {\alpha||\sigma\beta}{+~~~}}=A_\mu \hat{B}^\mu_{.\alpha\beta\sigma}-A_\alpha|_{\epsilon}N^\epsilon_{.\beta\sigma}-A_{\stackrel {\alpha||\mu}{+~~}}\hat{\Lambda}^\mu_{.\beta\sigma}
,
\end{equation}

\begin{equation}\label{32}
A_{\stackrel {\alpha||\beta\sigma}{-~~~}}-A_{\stackrel {\alpha||\sigma\beta}{-~~~}}=A_\mu \widetilde{\hat{B}}~^\mu_{.\alpha\beta\sigma}-A_\alpha|_{\epsilon}N^\epsilon_{.\beta\sigma}
-A_{\stackrel {\alpha||\mu}{-~~}}\hat{\Lambda}^\mu_{.\sigma\beta}
,
\end{equation}

\begin{equation}\label{33}
A_{\alpha||\beta\sigma}-A_{ \alpha||\sigma\beta}=A_\mu \overline{\hat{B}}~^\mu_{.\alpha\beta\sigma}
-A_\alpha|_{\epsilon}N^\epsilon_{.\beta\sigma},~~~~~~~~~~~~~~~
\end{equation}

\begin{equation}\label{36}
A_{\alpha\hat{;}\beta\sigma}-A_{ \alpha\hat{;}\sigma\beta}~=A_\mu \hat{R}^\mu_{.\alpha\beta\sigma}
-A_\alpha|_{\epsilon}N^\epsilon_{.\beta\sigma}~.~~~~~~~~~~~~~~~~
\end{equation}
Also, using the vertical derivative, we can write,
\begin{equation}\label{24}
A_\alpha|_{\beta\sigma}-A_\alpha|_{\sigma\beta}~=A_\mu \hat{S}^\mu_{.\alpha\beta\sigma},~~~~~~~~~~~~~~~~~~~~~~~~~~~~~~
\end{equation}
where,

\begin{equation}\label{19}
\hat{B}^\mu_{.\alpha\beta\sigma}\edf\hat{\Gamma}^{\mu}_{. \alpha \sigma\hat{,}\beta}-\hat{\Gamma}^{\mu}_{. \alpha\beta\hat{,}\sigma}+\hat{\Gamma}^{\mu}_{. \epsilon \beta}\hat{\Gamma}^{\epsilon}_{. \alpha\sigma}-\hat{\Gamma}^{\mu}_{. \epsilon \sigma}\hat{\Gamma}^{\epsilon}_{. \alpha\beta}-\hat{C}^\mu_{.\alpha\epsilon}N^\epsilon_{.\beta\sigma},~~~~~~~~~~~~~~~~
\end{equation}

\begin{equation}\label{34}
\widetilde{\hat{B}}~^\mu_{.\alpha\beta\sigma}\edf\widetilde{\hat{\Gamma}}~^{\mu}_{. \alpha \sigma\hat{,}\beta}-\widetilde{\hat{\Gamma}}~^{\mu}_{. \alpha\beta\hat{,}\sigma}+\widetilde{\hat{\Gamma}}~^{\mu}_{. \epsilon \beta}\widetilde{\hat{\Gamma}}~^{\epsilon}_{. \alpha\sigma}-\widetilde{\hat{\Gamma}}~^{\mu}_{. \epsilon \sigma}\widetilde{\hat{\Gamma}}~^{\epsilon}_{. \alpha\beta}-\hat{C}^\mu_{.\alpha\epsilon}N^\epsilon_{.\beta\sigma},~~~~~~~~~
\end{equation}

\begin{equation}\label{35}
\overline{\hat{B}}~^\mu_{.\alpha\beta\sigma}\edf\hat{\Gamma}^{\mu}_{. (\alpha \sigma)\hat{,}\beta}-\hat{\Gamma}^{\mu}_{. (\alpha\beta)\hat{,}\sigma}+\hat{\Gamma}^{\mu}_{. (\epsilon \beta)}\hat{\Gamma}^{\epsilon}_{. (\alpha\sigma)}-\hat{\Gamma}^{\mu}_{. (\epsilon \sigma)}\hat{\Gamma}^{\epsilon}_{. (\alpha\beta)}-\hat{C}^\mu_{.\alpha\epsilon}N^\epsilon_{.\beta\sigma},~~~~~
\end{equation}

\begin{equation}\label{37}
\hat{R}^\mu_{.\alpha\beta\sigma}\edf\hat{\cs{\alpha}{\sigma}{\mu}}_{\hat{,}\beta}-\hat{\cs{\alpha}{\beta}{\mu}}_{\hat{,}\sigma}+\hat{\cs{\epsilon}{\beta}{\mu}}\hat{\cs{\alpha}{\sigma}{\epsilon}} -\hat{\cs{\epsilon}{\sigma}{\mu}}\hat{\cs{\alpha}{\beta}{\epsilon}}
-\hat{C}^\mu_{.\alpha\epsilon}N^\epsilon_{.\beta\sigma},~
\end{equation}
and
\begin{equation}\label{21}
\hat{S}^\mu_{.\alpha\beta\sigma}\edf \hat{C}^{\mu}_{. \alpha \sigma:\beta}-\hat{C}^{\mu}_{. \alpha \beta:\sigma}+\hat{C}^{\mu}_{. \epsilon \beta}\hat{C}^{\epsilon}_{. \alpha \sigma}-\hat{C}^{\mu}_{. \epsilon \sigma}\hat{C}^{\epsilon}_{. \alpha\beta}.~~~~~~~~~~~~~~~~~~~~~~~~~~~~~~
\end{equation}\\

It can be shown; applying  the {\it Quotient Theorem} on the terms containing $N^\alpha_{.\mu\nu}
$,  appearing in (\ref{23}), (\ref{32}), (\ref{33}), (\ref{36}) and defined by (\ref{40}); that $N^\alpha_{.\mu\nu}$ is a third order tensor, skew symmetric in its lower indices. The geometric quantities
defined by  (\ref{19}), (\ref{34}), (\ref{35}), (\ref{37}) and (\ref{21}) are curvature tensors of the d-connections of the FAP-structure, respectively.\\

{\bf Theorem 2.} {\it "The horizontal curvature (\ref{19}) vanishes identically"} i.e. $\hat{B}^\mu_{.\alpha\beta\sigma}\equiv0$.

This theorem can be proved using definitions (\ref{9}) and (\ref{14}). \\

{\bf Theorem 3.} {\it "The vertical curvature (\ref{21}) vanishes identically} i.e. $\hat{S}^\mu_{.\alpha\beta\sigma}\equiv0$.

This theorem can be easily proved using definition (\ref{6}).\\

The curvature tensor $\hat{B}^\mu_{.\alpha\beta\sigma}$ (\ref{19}) can be expressed in terms of the  contortion tensor (\ref{0}) and curvature (\ref{37}) i.e.
\begin{equation}\label{49}
\hat{B}^\mu_{.\alpha\beta\sigma}=\hat{R}^\mu_{.\alpha\beta\sigma}+\hat{Q}^\mu_{.\alpha\beta\sigma}
\end{equation}where,
\begin{equation}\label{50}\hat{Q}^\mu_{.\alpha\beta\sigma}\edf\hat{\gamma}^\mu_{\stackrel{.\alpha\sigma||\beta}{~~~~+}}-\hat{\gamma}{ \stackrel{\mu}{+}}_{\stackrel{.\alpha\beta||\sigma}{+-~}}-\hat{\gamma}^\nu_{.\alpha\sigma}\hat{\gamma}^\mu_{.\nu\beta}
  +\hat{\gamma}^\nu_{.\alpha\beta}\hat{\gamma}^\mu_{. \nu \sigma},\end{equation}
  and $ \hat{R}^\mu_{.\alpha\beta\sigma}$ is defined by (\ref{37}). Now from the definitions (\ref{50}) and (\ref{37}), it is clear that neither $\hat{Q}^\mu_{.\alpha\beta\sigma}$ nor $ \hat{R}^\mu_{.\alpha\beta\sigma}$ vanishes, while their sum, given by (\ref{49}) vanishes (Theorem 2). It is to be considered that, the curvature tensor $\hat{R}^\mu_{.\alpha\beta\sigma}$ is  made of the linear connection $\hat{\cs{\alpha}{\beta}{\mu}} $ and the tensors $\hat{C}^{\mu}_{. \alpha\beta}$ and  $N^{\mu}_{. \alpha\beta}$ while the tensor $\hat{Q}^\mu_{.\alpha\beta\sigma}$ is purely made of the torsion (or the contortion) of the canonical connection. The addition of these two tensors gives rise to the vanishing of the total curvature of the space (Theorem 2). For the above mentioned comments, we call $\hat{Q}^\mu_{.\alpha\beta\sigma}$ the \underline{\textit{"additive inverse of the curvature tensor"}} \cite{WI} or the \textit{\underline{"anti-curvature"}} tensor \cite{W4}.\\

The following Table gives the possible contractions of the curvature tensors $\widetilde{\hat{B}}~^\mu_{.\alpha\beta\sigma}$ (\ref{34}), $\overline{\hat{B}}~^\mu_{.\alpha\beta\sigma}$ (\ref{35}) and $\hat{R}^\mu_{.\alpha\beta\sigma}$ (\ref{37}) in terms of the second order  tensors of Table 1. Note for example: $\widetilde{\hat{K}}_{\beta\sigma}\edf\widetilde{\hat{B}}~^\epsilon_{.\epsilon\beta\sigma}$
and $\widetilde{\hat{B}}_{\alpha\beta}\edf\widetilde{\hat{B}}~^\epsilon_{.\alpha\beta\epsilon}$.

\begin{center}
{\bf Table 2: Second Order contracted curvature Tensors}\\[0.2 cm]
\begin{tabular}{|l|l|}
\hline   &\\
 Skew Tensors& Symmetric Tensors\\[0.2 cm] \hline \hline&\\
  $\widetilde{\hat{K}}_{\beta\sigma}\edf\hat{\varepsilon}_{\sigma\beta}+\hat{\eta}_{\sigma\beta}$ & \\[0.2 cm]
  $\overline{\hat{K}}_{\beta\sigma}\edf\frac{1}{2}(\hat{\varepsilon}_{\sigma\beta}+\hat{\eta}_{\sigma\beta})$
  &\\[0.2 cm]$\hat{K}_{\beta\sigma}\equiv0$ & \\[0.2cm]\hline &\\[0.2 cm]
  $\widetilde{\hat{B}}_{[\alpha\beta]}\edf\hat{\chi}_{\alpha\beta}-\hat{\eta}_{\alpha\beta}-\frac{1}{2}\hat{\varepsilon}_{\alpha\beta}$&$\widetilde{\hat{B}}_{(\alpha\beta)}\edf-\frac{1}{2}~\hat{\theta}_{\alpha\beta}$\\[0.2 cm]\hline &\\
  $\overline{\hat{B}}_{[\alpha\beta]}\edf\frac{1}{2}\hat{\chi}_{\alpha\beta}-\frac{1}{4}(\hat{\eta}_{\alpha\beta}+\hat{\varepsilon}_{\alpha\beta})$&$\overline{\hat{B}}_{(\alpha\beta)}\edf-\frac{1}{4}~(\hat{\theta}_{\alpha\beta}-\hat{\sigma}_{\alpha\beta}-\hat{\omega}_{\alpha\beta}-\hat{\varpi}_{\alpha\beta})$\\[0.2 cm]\hline &\\
$\hat{R}_{[\alpha\beta]}\edf\frac{1}{2}(\hat{\chi }_{\alpha\beta}-\hat{\eta }_{\alpha\beta}-\hat{\varepsilon}_{\alpha\beta})$ &$\hat{R}_{(\alpha\beta)}\edf\frac{1}{2}(\hat{\psi }_{\alpha\beta}-\hat{\theta}_{\alpha\beta}-\hat{\phi}_{\alpha\beta})+\hat{\omega}_{\alpha\beta}$ \\[0.2 cm]\hline
\end{tabular}
\end{center}

It obvious that, the scalars of the contracted curvature tensors are made of the scalars
of the second order tensors (Sec. 3).\\
{  \bf Identities}\\
For d-connections $\hat{D}$, $\widetilde{D}$, $\overline{D}$ and $D$, the first and second Bianchi identities are, respectively\\
\begin{tabular}{ll}
First, & $\mathfrak{S}_{\mu,\nu,\sigma} (\hat{\Lambda}^{\alpha}_{\stackrel{. \mu \nu||\sigma}{ ~~~~+}}+\hat{\Lambda}^{\alpha}_{.\mu\epsilon}\hat{\Lambda}^{\epsilon}_{.\nu\sigma}+\hat{C}^{\alpha}_{. \mu \epsilon}N^{\epsilon}_{~\nu\sigma})\equiv0$.\\

& $\mathfrak{S}_{\mu,\nu,\sigma} (\hat{\Lambda}^{\alpha}_{\stackrel{.  \nu\mu||\sigma}{ ~~~~-}}+\hat{\Lambda}^{\alpha}_{.\mu\epsilon}\hat{\Lambda}^{\epsilon}_{.\nu\sigma}+\hat{C}^{\alpha}_{. \mu \epsilon}N^{\epsilon}_{~\nu\sigma})\equiv \mathfrak{S}_{\mu,\nu,\sigma}\widetilde{\hat{B}}~^\alpha_{.\mu\sigma\nu}$.\\
& $\mathfrak{S}_{\mu,\nu,\sigma} \hat{C}^{\alpha}_{. \mu \epsilon}N^{\epsilon}_{~\nu\sigma}\equiv \mathfrak{S}_{\mu,\nu,\sigma}\overline{\hat{B}}~^\alpha_{.\mu\sigma\nu}$.\\

& $\mathfrak{S}_{\mu,\nu,\sigma} \hat{C}^{\alpha}_{. \mu \epsilon}N^{\epsilon}_{~\nu\sigma}\equiv \mathfrak{S}_{\mu,\nu,\sigma}\hat{R}^\alpha_{.\mu\sigma\nu}$.\\ \\

Second, & $0~~~~\equiv
\mathfrak{S}_{\beta,\nu,\sigma}{ N^\epsilon_{~\nu\sigma}(\hat{\Gamma}^\alpha_{.\mu\beta:\epsilon}-\hat{C}^\alpha_{\stackrel{.\mu\epsilon||\beta}{~~~+}} +\hat{C}^\alpha_{.\mu \rho}[N^\rho_{~\beta:\epsilon}-\hat{\Gamma}^\rho_{.\epsilon\beta}])}$\\

 &$\mathfrak{S}_{\beta,\nu,\sigma}\widetilde{\hat{B}}~^\alpha_{\stackrel{.\mu\nu\sigma||\beta}{~~~~~-}}\equiv
\mathfrak{S}_{\beta,\nu,\sigma}\{\widetilde{\hat{B}}~^\alpha_{{.\mu\epsilon\sigma}} \hat{\Lambda}^\epsilon_{.\nu\beta} +N^\epsilon_{~\nu\sigma}(\widetilde{\hat \Gamma}~^\alpha_{.\mu\beta:\epsilon}-\hat{C}^\alpha_{\stackrel{.\mu\epsilon||\beta}{~~~-}} +\hat{C}^\alpha_{.\mu \rho}[N^\rho_{~\beta:\epsilon}-\widetilde{\hat \Gamma}~^\rho_{.\epsilon\beta}])\}$\\

 &$\mathfrak{S}_{\beta,\nu,\sigma}\overline{\hat{B}}~^\alpha_{.\mu\nu\sigma||\beta}\equiv
\mathfrak{S}_{\beta,\nu,\sigma}{N^\epsilon_{~\nu\sigma}(\hat{\Gamma}^\alpha_{.(\mu\beta):\epsilon}-\hat{C}^\alpha_{.\mu\epsilon||\beta} +\hat{C}^\alpha_{.\mu \rho}[N^\rho_{~\beta:\epsilon}-\hat{\Gamma}^\rho_{.(\epsilon\beta)}])}$\\

 &$\mathfrak{S}_{\beta,\nu,\sigma}\hat{R}~^\alpha_{.\mu\nu\sigma\hat{;}\beta}\equiv
\mathfrak{S}_{\beta,\nu,\sigma}{N^\epsilon_{~\nu\sigma}(\hat{\cs{\mu}{\beta}{\alpha}}_{:\epsilon}-\hat{C}^\alpha_{.\mu\epsilon\hat{;}\beta} +\hat{C}^\alpha_{.\mu \rho}[N^\rho_{~\beta:\epsilon}-\hat{\cs{\epsilon}{\beta}{\rho}}])}$\\

\end{tabular}
\section{The W-Tensor}

In constructing field theories, a variational method is usually applied to a Lagrangian function having certain properties.
In the case of geometric field theories, especially for gravity theories, the Lagrangian usually used is the
scalar curvature  $R$, or a general function $f(R)$ depending mainly on the curvature tensor (cf. Ref. 17).  In the case of geometries
with non-vanishing torsion $(T)$ some authors (cf. Ref. 18, 19) attempted to explore the role of using $f(T)$, in place of
$f(R)$, seeking a solution for problems facing gravity theories. In our point of view, it is better
to use a Lagrangian, function of both curvature and torsion, as each object plays a role, different
from that of the other, with possible mutual interaction.\\

In most known geometries, curvature and torsion are two distinct entities distinguished by two different
tensors. An important  object that reflects both curvature and torsion properties is the W-tensor \cite{WT}.
It can be considered as an amalgam made of both curvature and torsion. This tensor has been first defined \cite{W0}
 and used to construct a unified field theory \cite{W2}. It can be constructed,
only, in an AP-structure or one of its modifications. In what follows, we are going to construct
different forms of this tensor, in the FAP-space in order to make the geometry ready
for applications. Let us write the following commutation relations, corresponding to different H- and V- derivatives,

\begin{equation}\label{}
\h{i}_{\stackrel{\mu||\alpha\beta}{+~~~}}-\h{i}_{\stackrel{\mu||\beta\alpha}{+~~~}}= \h{i}_\epsilon  \hat{W}~^\epsilon_{.\mu\alpha\beta},\end{equation}
\begin{equation}\label{}\h{i}_{\stackrel{\mu||\alpha\beta}{-~~~}}-\h{i}_{\stackrel{\mu||\beta\alpha}{-~~~}}= \h{i}_\epsilon  \widetilde{\hat W}~^\epsilon_{.\mu\alpha\beta},\end{equation}
\begin{equation}\label{}\h{i}_{\mu||\alpha\beta}-\h{i}_{\mu||\beta\alpha}= \h{i}_\epsilon  \overline{\hat W}~^\epsilon_{.\mu\alpha\beta},\end{equation}
\begin{equation}\label{}\h{i}_{\mu\hat{;}\alpha\beta}-\h{i}_{\mu\hat{;}\beta\alpha}= \h{i}_\epsilon  \stackrel o {\hat W}~^\epsilon_{.\mu\alpha\beta},\end{equation}
\begin{equation}\label{}\h{i}_{\mu}|_{\alpha\beta}-\h{i}_{\mu}|_{\beta\alpha}= \h{i}_\epsilon  \stackrel \diamond {\hat W}~^\epsilon_{.\mu\alpha\beta},\end{equation}
where, when multiplying both sides of the above equations by $\h{i}^\sigma$, we obtain the following definitions of the W-tensor in the FAP-structure, respectively,

\begin{equation}\label{054} {\hat W}~^\sigma_{.\mu\alpha\beta}\edf\h{i}^\sigma ( \h{i}_{\stackrel{\mu||\alpha\beta}{+~~~}}-\h{i}_{\stackrel{\mu||\beta\alpha}{+~~~}})  \end{equation}

\begin{equation}\label{055}  \widetilde{\hat W}~^\sigma_{.\mu\alpha\beta}\edf\h{i}^\sigma( \h{i}_{\stackrel{\mu||\alpha\beta}{-~~~}}-\h{i}_{\stackrel{\mu||\beta\alpha}{-~~~}})\end{equation}

\begin{equation}\label{47}  \overline{\hat W}~^\sigma_{.\mu\alpha\beta}\edf\h{i}^\sigma ( \h{i}_{{\mu||\alpha\beta}}-\h{i}_{{\mu||\beta\alpha}})\end{equation}

\begin{equation}\label{48}  \stackrel o {\hat W}~^\sigma_{.\mu\alpha\beta}\edf\h{i}^\sigma ( \h{i}_{{\mu\hat{;}\alpha\beta}}-\h{i}_{{\mu\hat{;}\beta\alpha}})\end{equation}

\begin{equation}\label{} \stackrel \diamond{\hat W}~^\sigma_{.\mu\alpha\beta}\edf\h{i}^\sigma (
\h{i}_{\mu}|_{\alpha\beta}-\h{i}_{\mu}|_{\beta\alpha})\end{equation}

{\bf Theorem 4.}
{\it The curvature and W-tensors have the following coincidences:
  \begin{description}
   \item[(i)]  $ \overline{\hat W}~^\sigma_{.\mu\alpha\beta}$ coincides with the curvature tensor  $\overline{\hat B}~^\sigma_{.\mu\alpha\beta}$.
   \item[(ii)] $\stackrel o {\hat W}~^\sigma_{.\mu\alpha\beta} $ coincides with the curvature tensor $\hat R^\sigma_{.\mu\alpha\beta}$.
 \end{description}}
\underline{Proof.}

\begin{description}
  \item[(i)]
  Using (\ref{31}), we can write (\ref{47}) in the form
  \begin{equation}\label{047}  \overline{\hat W}~^\sigma_{.\mu\alpha\beta}\edf \frac{1}{2}(\hat{\Lambda}^\sigma_{.\mu\alpha||\beta}-\hat{\Lambda}^\sigma_{.\mu\beta||\alpha})+\frac{1}{4}(\hat{\Lambda}^\nu_{.\mu\beta}\hat{\Lambda}^\sigma_{.\alpha \nu}-\hat{\Lambda}^\nu_{.\mu\alpha}\hat{\Lambda}^\sigma_{.\beta \nu}),\end{equation}
  since it can easily shown that,
  $$ \hat{\Lambda}^\alpha_{.\mu\nu||\sigma}=\hat{\Lambda}^\alpha_{\stackrel{.\mu\nu||\sigma}{~~~+}}+\frac{1}{2}
  \mathfrak{S}_{\mu,\nu,\sigma}\hat{\Lambda}^\epsilon_{.\mu\nu}\hat{\Lambda}^\alpha_{.\sigma\epsilon}.$$
Then substituting into (\ref{047}), (i) follows.
  \item[(ii)] Using (\ref{-2}), we can write (\ref{48}) in the form,
\begin{equation}\label{048}  \stackrel o {\hat W}~^\sigma_{.\mu\alpha\beta}\edf \hat{\gamma}^\sigma_{\stackrel{.\mu\alpha||\beta}{~~~~+}}-\hat{\gamma}^\sigma_{\stackrel{.\mu\beta||\alpha}{~~~~+}}+\hat{\gamma}^\nu_{.\mu\beta}\hat{\gamma}^\sigma_{.\nu\alpha}
 -\hat{\gamma}^\nu_{.\mu\alpha}\hat{\gamma}^\sigma_{. \nu \beta}+\hat{\gamma}^\sigma_{. \mu \nu} \hat{\Lambda}^\nu_{.\alpha \beta},\end{equation}
  but we have the relation,
  $$\hat{\gamma}^\alpha_{\stackrel{.\mu\nu||\sigma}{~~~~+}}=
      \hat{\gamma}{ \stackrel\alpha{ +}}_{\stackrel{.\mu\nu||\sigma}{+-~}}+\hat{\gamma}^\alpha_{.\mu\epsilon}\hat{\Lambda}^\epsilon_{.\sigma\nu},$$
which by substitution into (\ref{048}) proves the second part of the theorem.

\end{description}

As a consequence of identities (\ref{13}) and (\ref{14}) , we get ${\hat W}~^\sigma_{.\mu\alpha\beta}\equiv0$ and $\stackrel \diamond{\hat W}~^\sigma_{.\mu\alpha\beta}\equiv0$.

\section{Discussion and Concluding Remarks}
\begin{description}
\item[(1)] The structure given in this article shows that all geometric objects, including the non-linear connection(\ref{4}), are define from the building blocks of the FAP-structure i.e. the vector fields $\h{i}_\mu$.
\item[(2)] The torsion corresponding to the non-linear connection $(N^\alpha_{~\beta:\sigma}-N^\alpha_{~\sigma:\beta})$ coincides with the torsion of the Bervald type linear connection $\hat{T}^\alpha_{.
\mu\nu}$ given by (\ref{42}).
  \item[(3)] The anti-curvature tensor defined in an AP-structure has been shown
to be useful in describing antigravity \cite{W4}. It is made of the torsion
of the AP-structure. The definition
of this tensor, in the FAP-structure, shows that it contains, implicity,
the non-linear connection. Applications, using FAP will illuminate
the role of the non-linear connection in this respect, if any.
  \item[(4)] Its W-tensor (Sec. 5) reflects, not only the effects of the curvature
and torsion (as in the case of AP-structure \cite{WT}), but also the effect
of the non-linear connection. This is an important
property which enables us to explore the role of non-linear connections,
in unifying fundamental interactions, if any.

\item[(5)]
The following table (Table 3) gives a brief comparison
between the FAP-, the AP-,  Finsler  and  Riemannian
structures.\newpage
\begin{center}{\bf Table 3: Comparison between FAP-, AP-, Finsler and Riemann  geometries}\\[0.2cm]

\begin{tabular}{|l|c|c|c|c|}
 \hline 
Criteria&FAP&AP&Finsler&Riemann\\\hline\hline 
&&&&\\ Underlying Space&$(M,\E{i})$&$ (M,\q{i})$&$(M,F) $&$ (M,g)$\\ &&&&\\\hline
&&&&\\Structure Dependence&$\E{i}(x,y)$&$ \q{i}(x)$&$F(x,y) $&$ g(x)$\\ &&&&\\\hline
&&&&\\Number of Natural Metric  &$2$&$2$&$1$&$1$\\
Connections&&&&\\ &&&&\\\hline
&&&&\\Number of built-in Non-Linear &$2$&0&$1$&0\\
Connections&&&&\\&&&&\\ \hline
&&&&\\Number of Anti-curvature  &1&1&0&0 \\
Tensors&&&&\\ &&&&\\\hline
&&&&\\Number of W-Tensors* 
&1&1&0&0\\&&&&\\ \hline
\end{tabular}
\end{center}
*\textit{Non-Vanishing and differs from the Curvature tensor.}\\\\
From this table, it is clear that the FAP-structure is more
thorough than both conventional AP and Finsler structures, in
particular:

\begin{description}
   \item[(i)]  It is clear that the only non-vanishing W-tensor
and different from the curvature tensors is that given by (\ref{055}). Using  (\ref{30}) this tensor can be written in the form
\begin{equation}\label{0055}  \widetilde{\hat W}~^\sigma_{.\mu\alpha\beta}\edf \hat{\Lambda}^\sigma_{\stackrel{.\mu\alpha||\beta}{~~~~-}}-\hat{\Lambda}^\sigma_{\stackrel{.\mu\beta||\alpha}{~~~~-}}+\hat{\Lambda}^\nu_{.\mu\beta}\hat{\Lambda}^\sigma_{.\alpha \nu}-\hat{\Lambda}^\nu_{.\mu\alpha}\hat{\Lambda}^\sigma_{.\beta \nu}.\end{equation}
Due to the importance of this tensor in physical applications, we are going to write its possible contractions which will be written, in terms of the tensors of Table 1, as
\begin{equation}\label{00}
\widetilde{\hat S}_{\alpha\beta}\edf \widetilde{\hat W}~^\epsilon_{.\epsilon\alpha\beta} = \hat{\varepsilon}_{\beta\alpha}+2\hat{\eta}_{\beta\alpha} ,\end{equation}
\begin{equation}\label{01}\widetilde{\hat W}_{.\mu\alpha}\edf  \widetilde{\hat W}~^\epsilon_{.\mu\alpha\epsilon} =\hat{\chi}_{\mu\alpha}-\hat{\eta}_{\mu\alpha}-\hat{c}_{\stackrel{\mu||\alpha}{~~+}}
-\hat{\varpi}_{\alpha\mu}+\hat{\sigma}_{\alpha\mu}+\hat{\omega}_{\alpha\mu}  \end{equation}
The tensor (\ref{00}) is anti-symmetric. So, it cannot be further contracted. The tensor (\ref{01}) is non-symmetric. So, its symmetric part can be contracted to give the scalar
\begin{equation}\widetilde{\hat W} =  -\frac{1}{2} \hat{\theta}+3\hat{\omega} +\hat{\sigma} \end{equation}
  \item[(ii)] We stress again that, the W-tensor, defined in Sec. 5 contains the effects of the curvature tensor, the torsion of the linear connection, and the non-linear connection. It is to be considered that this tensor has only one value given by (\ref{055}). Other values (\ref{054}), (\ref{47}), (\ref{48}) are either vanishing (due to (\ref{13}) and (\ref{14})) or coincides with the curvature tensors (Theorem 4). This is an important property, of the FAP-structure, which would lead to the construction of a unique field theory using this tensor.
\end{description}

\end{description}

The following  schematic diagram shows the relations
between the above mentioned geometric structures. Solid arrows
represent the direction towards special cases, while dashed arrows show
the associated structures.\\
\begin{center}
{
\begin{tabular}{ccc}
\fcolorbox{black}{white}{~FAP~}&$\xrightarrow{\rm \hat{C}_{\mu\nu\sigma}~=~0,~({ \it Theorem 1})}$&\fcolorbox{black}{white}{~~~~AP~~~} \\
\rotatebox{-90}{$ \stackrel{\hat{C}_{[\mu\nu]\sigma}=0,~{\it \cite{W1}}}{------\dashrightarrow}$}& &\rotatebox{-90}{$ \stackrel{{g}_{\mu \nu} =  \q{i}_{\mu} \q{i}_{\nu},~({\it cf. \cite{W3}})}{------\dashrightarrow}$}\\
\fcolorbox{black}{white}{Finsler}&$\xrightarrow{\rm ~~~C_{\mu\nu\sigma}~=~0,~({\it cf. \cite{F1}})~~~}$&\fcolorbox{black}{white}{Riemann}
\end{tabular}
}\\
\end{center}
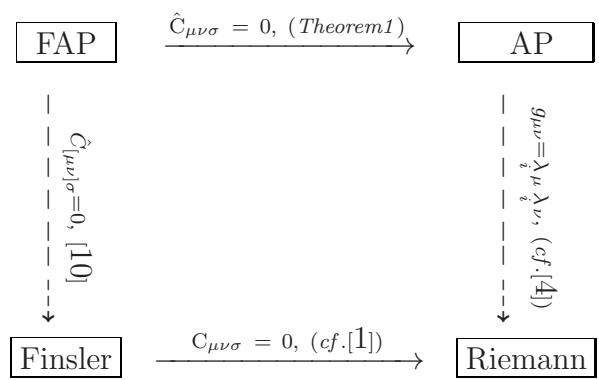
\begin{center}Fig. 1: Relation between different geometric structures. \end{center}


\end{document}